\documentclass[vecphys]{svmult}

\usepackage{makeidx}         
\usepackage{graphicx}        
\usepackage{multicol}        
\usepackage[bottom]{footmisc}

\begin{document}

\title*{The Local Velocity Anomaly}
\author{R. Brent Tully}
\institute{Institute for Astronomy, University of Hawaii, Honolulu, USA}

\maketitle

\section{Historical Notes} 
\label{sec:1}
\subsection{What is the Local Velocity Anomaly?}

\begin{figure}
\begin{center}
\includegraphics[scale=0.56]{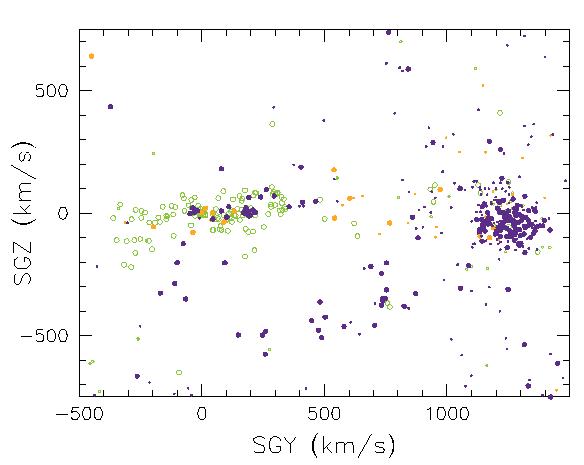}
\caption{Peculiar velocities $<-100$~km/s are coded solid purple (black),
peculiar velocities $>+100$~km/s are coded solid orange (grey), and
peculiar velocities within 100~km/s of zero are coded green (grey) with 
open symbols.  Our Galaxy lies at SGY=SGZ=0 within the ``Local Sheet''.
The Leo Spur lies at negative SGZ and the Virgo Cluster is the dense 
clump of galaxies at the right.}
\label{ls}
\end{center}
\end{figure}

The Local Velocity Anomaly was first discussed at the Vatican Conference
in 1988 \cite{fab88,tul88} and was given attention for a few years
\cite{gir90, han90, tul92}.
Subsequently, there has been little notice of the phenomenon.  The 
observation is the following: our Galaxy lies in a structure that has
very low internal motions in a cosmic expansion reference frame but 
there is a discontinuity of velocities as we step to the nearest
adjacent structures.  
A modern view of peculiar velocities is provided by Figure~\ref{ls} 
\cite{tul07}. 
In earlier discussions, we referred to our home
structure as the Coma--Sculptor Cloud.  Here we refer to the 
co-moving part of it as the `Local Sheet'. 
The nearest adjacent 
structure is the Leo Spur, found to have large negative peculiar velocities.

There is a relationship to the historic Hubble Constant controversy.
There was a point of view \cite{bot86,kra86} that what was seen as the
Local Velocity Anomaly was just a manifestation of Malmquist bias. 
The normal consequence of Malmquist bias is underestimation
of distances that gets worse as one goes farther away, resulting in increasing 
estimates of the Hubble parameter with distance \cite{san99}.  An upturn
in the Hubble parameter is seen at $\sim 1000$~km~s$^{-1}$.  Is it due
to Malmquist bias (and low H$_0$) or is it something that, at least in 
part, requires 
a physical explanation?  Malmquist bias can be mitigated \cite{tul00}.
In any event, modern data shows unambiguously that the
Local Velocity Anomaly is a real phenomenon.

\subsection{The Great Attractor Problem}

\begin{figure}
\begin{center}
\includegraphics[scale=0.66]{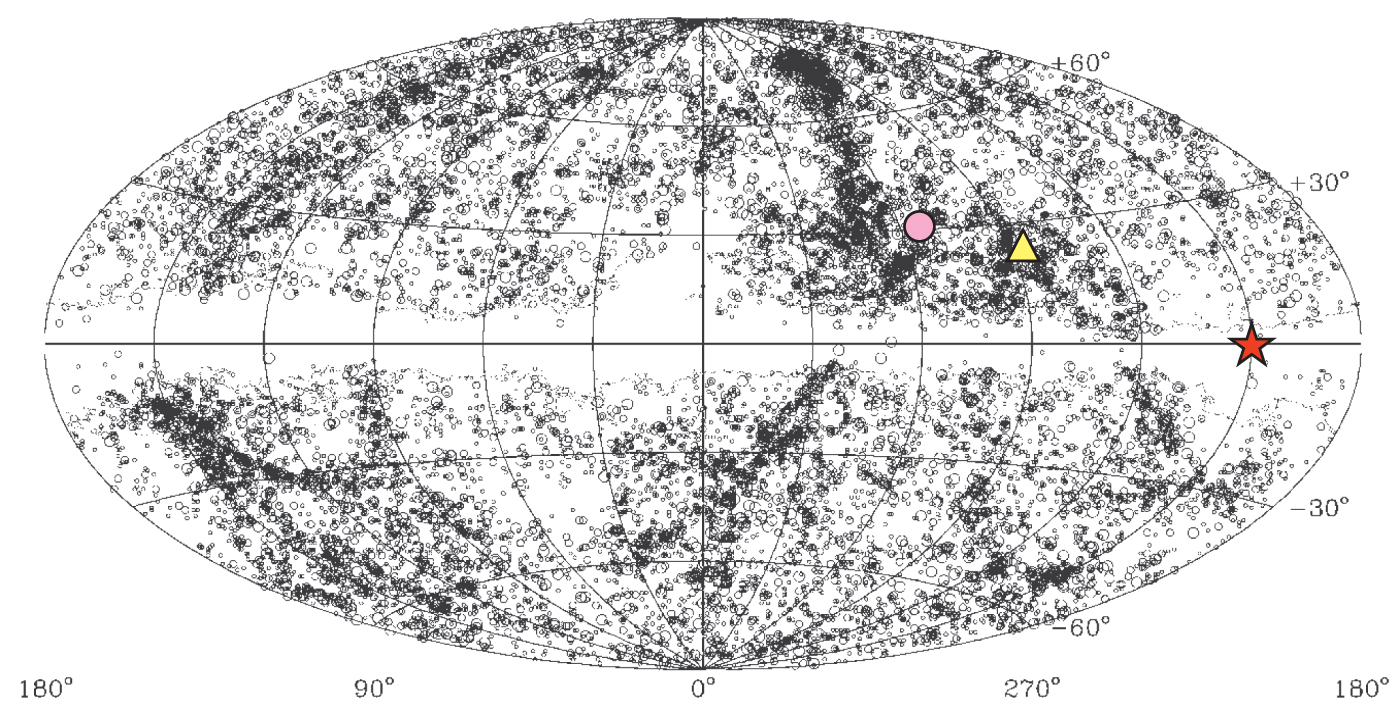}
\caption{Distribution of galaxies in Galactic coordinates, adapted from
reference \cite{kra00}.  The CMB dipole apex is indicated by the yellow
triangle.  The dipole in the distribution of observed galaxies is 
indicated by the pink circle.  The offset could be due to sources
in the plane of the Milky Way, say centered in the vicinity of the red star.
Alternatively, the offset might be attributed to a {\it lack} of galaxies
$180^{\circ}$ removed in longitude.
}
\label{lahav}
\end{center}
\end{figure}

Since that same Vatican Conference there has been an appreciation that
what we see in the distribution of galaxies does not adequately explain 
the motion inferred by the
Cosmic Microwave Background (CMB) dipole \cite{lyn88}.
Our motion with respect to the CMB reference frame is close to the plane
of our Galaxy (even after subtraction of our Galactic orbital motion).
The dipole in the distribution of galaxies, as determined back in 1988
from the distribution of galaxies seen in Figure~\ref{lahav} \cite{kra00} 
was near but not precisely aligned with the CMB
dipole direction.  It was reasonable to suspect that there were many
galaxies hidden in the zone of avoidance and a considerable industry
developed to look for them \cite{koc07, kra00}.
Objects were found but not in the quantity
required to explain the offset of the CMB dipole direction away from the 
supergalactic equator (defined by the ridgeline of the pronounce vertical
band to the right of center in the plot) toward the supergalactic south
(to the right of that ridgeline). Already in 1988 \cite{lyn88},
the prescient but unsubstantiated suggestion was made that this offset
might be the consequence of a push from the Local Void. 

\subsection{Dynamical Models}

There have been attempts to reconstruct the development
of local structure \cite{phe01,sha95}. Numerical Action models involve 
the reconstruction of orbits to 
match the observed distribution of galaxies.  Masses are assigned to
galaxies in proportion to light.  Resultant models are evaluated
through a comparison of predicted and observed peculiar velocity fields.
The greatest deficiency of the models has been an inability to explain
the {\it amplitude} of the motions seen in the Local Velocity Anomaly.
Model velocities were only large enough with unreasonably large masses 
assigned to nearby galaxies.  More on the resolution of this problem in 
the ensuing discussion.

\subsection{The Local Void}

The Local Void was identified in the {\it Nearby Galaxies Atlas} \cite{tul87}.
It begins at the edge of the Local Group.  It is poorly understood
because much of it lies behind the plane of the Milky Way.  It is hard to
see nothing, especially when it is obscured.  The HI Parkes All Sky Survey
\cite{mey04} penetrates the zone of avoidance and clearly shows that the
Local Void is to be taken seriously.  The Local Void occupies much of the
foreground in the underdense region in the left half of Fig.~\ref{lahav}. 

\section{What's New?}

We now have a database of almost 1800 distances to galaxies within 3000~km/s.
Over 600 of these have accuracies better than 10\%, derived from either
the Cepheid Period--Luminosity \cite{fre01}, Tip of the Red Giant Branch (TRGB)
\cite{sak96}, or Surface Brightness Fluctuation \cite{ton01} methods.
The rest are based on luminosity--linewidth measures \cite{kar02,tul77,tul00}.
With distances, $d$, it is possible to separate the radial component of
peculiar velocities, $V_{pec}$, from observed velocities, $V_{obs}$:
$V_{pec} = V_{obs} - {\rm H}_0 d$.  In this analysis we take H$_0=74$~km/s/Mpc.

\begin{figure}
\begin{center}
\includegraphics[scale=0.58]{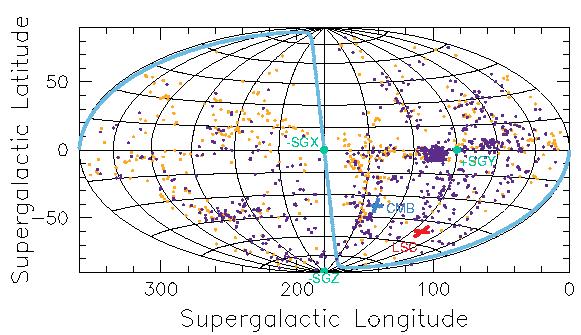}
\caption{Peculiar velocities of galaxies within 3000~km/s projected in
supergalactic coordinates.  Peculiar velocities $>+100$~km/s are indicated
by orange (grey) circles and peculiar velocities $<-100$~km/s are in purple
(black).  The Virgo Cluster is seen as the dense knot of objects near the
+SGY axis.  The motion of the Local Sheet with respect to these galaxies
is toward the red cross labeled `LSC' (Local Supercluster).  The direction
of the CMB vector in the same rest frame is indicated by the blue cross
labeled `CMB'.  The blue (grey) band indicates the plane of our Galaxy. 
The directions of the supergalactic --SGX, +SGY, and --SGZ axes are labeled.
}
\label{pv_aitoff}
\end{center}
\end{figure}

A map of peculiar velocities is shown in the aitoff projection in 
supergalactic coordinates of Figure~\ref{pv_aitoff}.  It is seen that motions 
in the lower right quadrant are overwhelmingly toward us while motions in the
upper left quadrant are predominantly away from us.  This pattern can be
explained if we have a velocity toward the lower right quadrant.  The 
inferred apex of our motion is identified on the figure by the cross
labeled `LSC'.  The relationship with the CMB dipole should be
noted.

\begin{figure}
\begin{center}
\includegraphics[scale=0.58]{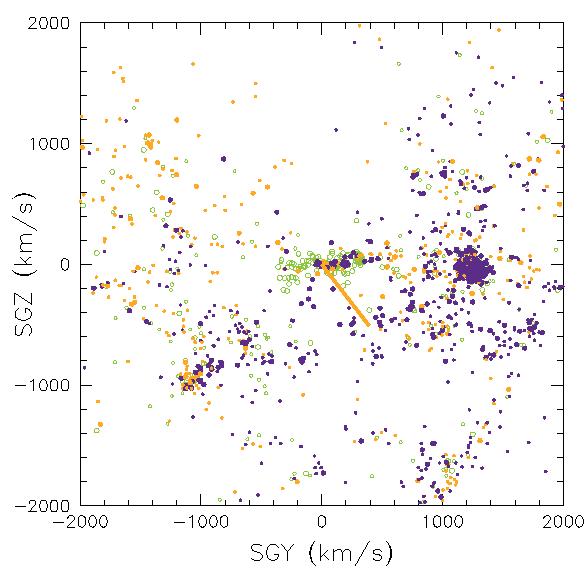}
\caption{Peculiar velocities seen by an observer on the +SGX axis in 
supergalactic coordinates.  Large negative and positive peculiar velocities
are shown in orange (grey) and purple (black) as in the previous figure.
As in Fig.~\ref{ls}, galaxies with peculiar velocities within 100 km/s 
of zero are shown by open green (grey) circles.  The Local Sheet, the 
horizontal structure around the origin, is moving in the direction of the
orange (grey) bar toward the lower right. The Virgo Cluster is at
SGY=1200~km/s, SGZ=0~km/s; the Fornax Cluster is at SGY=--1000~km/s,
SGZ=--1000~km/s.
}
\label{pv_far}
\end{center}
\end{figure}

Figure~\ref{pv_far} uses the same data but steps from our location at the 
center of
the scene to a vantage point looking in from a large distance.  From the
viewing position that is chosen, once again galaxies in the lower right tend 
to have negative peculiar velocities and galaxies in the upper left tend to
have positive peculiar velocities.  {\it The pattern is explained if we are
moving with respect to the ensemble of observed objects toward the lower 
right, in the direction of the bar emanating from the origin.}

\begin{figure}
\begin{center}
\includegraphics[scale=0.58]{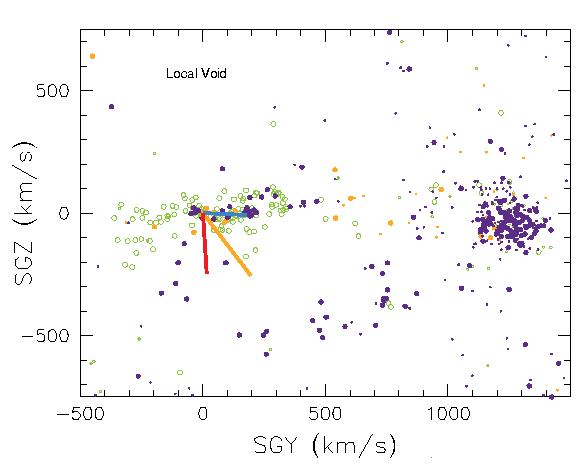}
\caption{Blow up of the central region of Fig.~\ref{pv_far}.  The vector
of the motion of the Local Sheet with respect to galaxies with measured
distances less than 3000~km/s is indicated by the orange (grey) bar pointing
toward the lower right.  The horizontal blue (black) bar is the component 
of the Local Sheet motion directed toward the Virgo Cluster, the clump at the
right edge of the figure.  The residual, after subtraction of the component
toward the Virgo Cluster from the observed Local Sheet motion, is the red
(black) bar directed almost straight down.  Motion in this direction is
suspected to be due to expansion away from the Local Void at positive SGZ.
}
\label{pv_near}
\end{center}
\end{figure}

We zoom in on this scene with Figure~\ref{pv_near}.  It is seen that the 
pattern of
peculiar velocities (negative to lower right, positive to upper left) only
begins beyond $\sim 7$~Mpc, outside our immediate structure.  Within 7~Mpc,
almost 200 galaxies now have well observed TRGB distances and are found
to be expanding with roughly the Hubble law with very low peculiar 
velocities \cite{kar03}.  {\it This entire local region is moving together
in the direction toward the lower right.}  We are calling our co-moving
flattened structure the {\it Local Sheet}.

The Virgo Cluster is the prominent knot of objects with net negative 
velocities at the right edge of Fig.~\ref{pv_near}.  It has long been
appreciated that this structure influences our motion \cite{aar82} and
dynamical models provide an estimate of the amplitude \cite{moh05}.
The velocity of the Local Sheet with respect to galaxies beyond 7~Mpc and 
within 3000~km/s is 323~km/s. A component is directed toward the Virgo Cluster
with an amplitude of 185~km/s, which is in reasonable accord with 
expectations given $\sim 1 \times 10^{15}~M_{\odot}$ in the cluster at
17~Mpc.  If this component is subtracted from the observed vector, the 
result is a vector of amplitude 259~km/s in a direction close to the 
supergalactic south pole.  This direction is orthogonal to the disk of
the Local Sheet.  It is not toward anything substantial, but it is 
directly {\it away} from the Local Void.

\begin{figure}
\begin{center}
\includegraphics[scale=0.53]{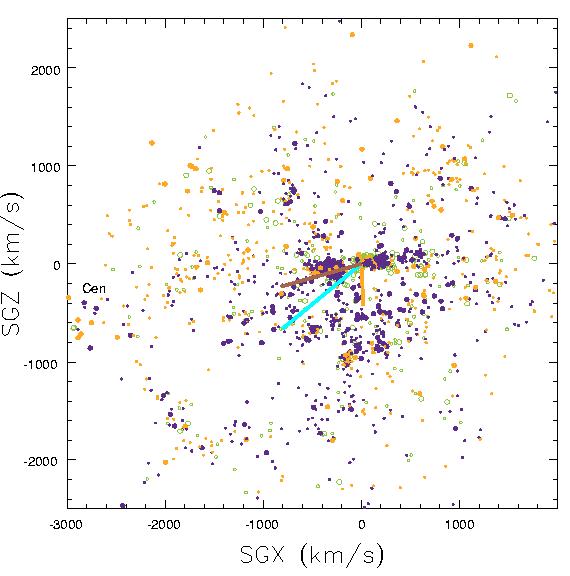}
\caption{Peculiar velocities seen by an observer looking in along the --SGY
axis.  From this angle, rotated 90 degrees from Fig.~\ref{pv_far}, the 
orange (grey) vector of the motion of the Local Sheet with respect to galaxies
within 3000~km/s is pointed almost straight down.  The CMB vector in the same
frame of rest is indicated by the cyan (grey) bar pointing down and to the 
left.  The difference, if the orange vector is subtracted from the cyan vector,
gives the brown (black) bar pointing almost horizontal to the right.
This is a vector attributed to influences on scales greater than 3000~km/s.
It is directed quite close to the position of the Centaurus Cluster at 
SGX=--2700~km/s, SGZ=--500km/s.
}
\label{pv_lss}
\end{center}
\end{figure}

We will return to a consideration of the motion away from the Local Void
in the next section, but let us first look at the importance of this
component in the inventory that makes up the motion reflected in the CMB
dipole.  With Fig.~\ref{pv_lss} we have rotated $90^{\circ}$ in the
supergalactic equatorial plane and see that the orange vector displayed 
in Fig.~\ref{pv_far} is directed almost straight down in this new view.
The cyan vector directed toward the lower left is the projection of the 
CMB dipole vector of 631~km/s.  The difference between these two is the
brown vector of 455~km/s, pointing almost horizontal (ie, close to the 
supergalactic equator) to the left.  This component was not picked up
by our sample of distances, so is attributed to structure at velocities
greater than 3000~km/s.

\section{Voids Push}

In a more complete description of this research \cite{tul07} it is shown 
that a completely empty hole in an otherwise uniform medium in a 
$\Lambda$CDM cosmology with matter density $\Omega_m = 0.3$ expands at
16~km/s/Mpc.  In the present situation, we want to interpret 260~km/s as
expansion away from the Local Void.  It is required that the Local Void
have a diameter of at least 45~Mpc.  To the degree that it is not empty, 
it must be larger.

\begin{figure}
\begin{center}
\includegraphics[scale=0.58]{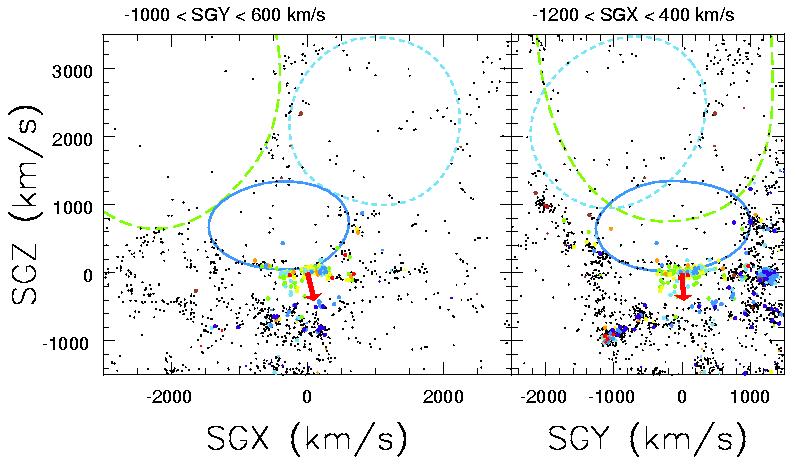}
\caption{Two projections of the region of the Local Void.  The ellipses
outline three apparent sectors of the Local Void.  The solid dark blue ellipse
shows the projection of the nearest part of the Local Void, bounded at our
location by the Local Sheet.  North and South extensions of the Local Void
are identified by the light blue short--dashed ellipse and the green 
long--dashed ellipse, respectively.  These sectors are separated by bridges
of wispy filaments.  The red vector indicates the direction and amplitude of 
our motion away from the void.
}
\label{void}
\end{center}
\end{figure}

There is an attempt to illustrate the region of the Local Void with 
Figure~\ref{void}.  It is difficult to give a fair representation, first
because much of it is concealed behind the center of our Galaxy, and
second because it is so very large.  The region is not entirely empty
of galaxies.  Several minor filaments lace through the volume and motivate
us to separate the nearest part of the Local Void from more distant
North and South extensions.  On observational grounds, an ensemble void
region of greater than 45~Mpc cannot be excluded.  

A probem mentioned earlier with dynamic models is now illuminated.
Those models require some assumption about the distribution of matter in
the zone of avoidance.  A priori, it would be unreasonable to assume that
this region is empty, and modelers use a variety of recipes that usually
result in assigning the mean density to unseen places.  Future models
should consider what happens if the sector of the Local Void is left empty. 

Perhaps the most interesting issue raised in this study is the following.  
There are 
debates about whether voids are really empty of matter or whether they are 
just low density regions where star formation is inefficient.  Researchers
have studied voids at large distances in attempts to answer this
question but have given little attention to the void that starts only 1~Mpc
away.  We are in a unique position to study the peculiar motion of the 
shell bounding our own void.  The observed motion away from the Local Void,
the Local Velocity Anomaly, might represent the best available evidence that
voids are really empty.

\vskip 1.5cm

This report summarizes research undertaken with Ed Shaya, Igor Karachentsev,
H\'el\`ene Courtois, Dale Kocevski, Luca Rizzi, and Alan Peel and published
as reference \cite{tul07}.  Videos in support of the discussion are found at
http://www.ifa.hawaii.edu/~tully/.  Financial support has been provided by
the US National Science Foundation and Space Telescope Science Institute.


%
%
%
%
%



\end{document}